\documentclass[twocolumn,prl,superscriptaddress,showpacs]{revtex4}
\usepackage{epsfig}
\usepackage{times}
\usepackage{amsmath}

\begin{document}

\title{Density Matrix Renormalization Group Study of Incompressible Fractional Quantum Hall States}

\author{A.~E.~Feiguin}
\affiliation{Microsoft Station Q, University of California, Santa Barbara, California 93106}
\author{E.~Rezayi}
\affiliation{Department of Physics, California State University, Los Angeles, California 90032 }
\author{C.~Nayak}
\affiliation{Microsoft Station Q, University of California, Santa Barbara, California 93106}
\author{S.~Das Sarma}
\affiliation{Condensed Matter Theory Center, Department of Physics, University of Maryland,
College Park, Maryland 20742}

\date{\today}

\begin{abstract}
We develop the Density Matrix Renormalization Group (DMRG) technique for numerically
studying incompressible fractional quantum Hall (FQH) states on the 
sphere. We calculate accurate estimates for ground state energies and excitation
gaps at FQH filling fractions $\nu=1/3$ and $\nu=5/2$ for systems that are almost
twice as large as the largest ever studied by exact diagonalization.
We establish, by carefully comparing with existing numerical results on smaller systems, that DMRG is a highly effective numerical tool for studying incompressible FQH states.

\end{abstract}

\pacs{73.43.Cd, 5.10.Cc}

\maketitle

\paragraph{Introduction --}
Numerical techniques, particularly the exact diagonalization of a small 
system of $N_e$ interacting 2D electrons in a strong perpendicular magnetic 
field, have been a central and essential tool in the theoretical study of the
fractional quantum Hall effect (FQHE) \cite{Tsui} dating back, rather curiously, 
to the days \cite{YHL,Laughlin1} before Laughlin wrote down \cite{Laughlin2} his famous 
wavefunction. This wavefunction has served as the paradigm for understanding the 
odd-denominator incompressible FQH states during the last 25 years
due, in part, to supporting evidence given by exact diagonalization
studies: the Laughlin wavefunction has almost unity overlap with the numerically 
calculated exact many-body wavefunction for small ($N_e = 4-8$) spherical 
systems interacting via the Coulomb interaction, as shown in the seminal work
by Haldane \cite{Haldane}.  Of equal importance, 
Haldane developed an extremely enlightening pseudopotential expansion in the 
conserved relative orbital angular momentum channel, 
establishing in the process the amazing fact that the Laughlin 
wavefunction is the exact eigenstate for the short-ranged interaction potential
${V_1}\neq 0$, ${V_J}=0$ for $J=3,5,\ldots$, where the $V_J$ is the interaction between
two electrons with relative angular momentum $J$.  This 
immediately led to a deep physical and mathematical understanding of why 
and when a Laughlin-type incompressible state would be a good 
description of the ground state observed in experiments. Ever since Haldane's 
work, exact diagonalization of finite systems, mostly on the sphere, torus, or disc,
has been a crucial theoretical tool for studying FQH systems -- indeed, the only
one with great predictive power for the dependence of the ground state
on material parameters.
It is therefore of substantial interest and importance that new numerical 
methods are developed for larger system sizes which 
could be useful in understanding more complex and exotic FQH states.
In particular, there is great current interest in higher Landau level (LL) FQH states in the 
first \cite{Xia} or even the second \cite{Gervais} LL and the fascinating even denominator 
states (e.g. the $5/2$ FQH state \cite{Willet} discovered 20 years ago) as well as 
studies of spin polarization properties \cite{Eisenstein} of FQH states. These all require 
finite size studies involving system sizes $N_e$ much larger than what has 
been achieved through the exact diagonalization method, which is typically 
restricted to $N_e = 10-18$ depending on the LL filling factor $\nu$. Since 
the Hilbert space dimension grows exponentially with $N_e$, the 
exact diagonalization technique has essentially reached its (memory and 
storage) limit, and no significant advance in increasing $N_e$ for exact 
diagonalization is anticipated in the near future.

In this Letter we use a powerful technique, the Density Matrix Renormalization
Group (DMRG) \cite{dmrg,uli}, for studying ground and excited state properties of incompressible FQH states, 
allowing a substantial jump ($\sim$ by almost a factor of 2!) in $N_e$
(up to $N_e = 20$ for $\nu = 1/3$  and 30 for $\nu = 5/2$).  
We introduce, establish, and 
validate this technique by concentrating on the archetypal FQH state at 
$\nu = 1/3$ and the mysterious and not yet well-understood FQH state at
$\nu = 5/2$. Our work not only reproduces the numerical 
results obtained earlier in exact diagonalization studies for the same 
values of $N_e$, but also extends our understanding of
the ground state and low-lying excited state properties of these FQH states
to substantially larger system sizes. Our treatment of large systems on the sphere depends on the use of
intricate, non-trivial methods which are described below, unlike previous pioneering applications of the DMRG to the quantum Hall
problem, which concentrated mostly on compressible non-quantized states (e.g. bubbles and stripes) in higher Landau levels
on the torus \cite{Shibata}. We suggest that these techniques will allow, in the 
future, DMRG studies of problems simply impossible to treat with exact 
diagonalization. We believe that the development 
of the DMRG method (which is almost exclusively used for 1D interacting 
systems) to the 2D FQHE problem should have high 
impact on both the physics of the FQHE and the theory of the DMRG.

\begin{centering}
\begin{figure}
\epsfig {file=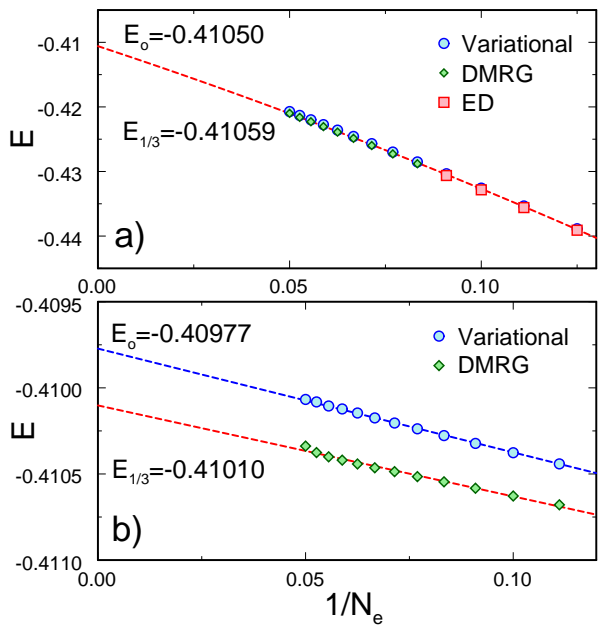,width=80mm}
\caption{
a) Ground state and variational energies at $\nu=1/3$ obtained with DMRG and exact diagonalization (ED). The extrapolated values are obtained using a quadratic function of $1/N_e$. Variational results are barely visible because they lie on top of the DMRG energies at this scale.
b) Finite size scaling using a renormalized magnetic length $\ell_0$, following Refs.\cite{Morf, Morf and Shankar}.
}
\label{fig1}
\end{figure}
\end{centering}

\paragraph{DMRG --}
The Density Matrix Renormalization Group (DMRG) \cite{dmrg,uli}
method is a sophisticated algorithm to 
calculate the ground-state and low-lying excitations of
quantum many-body problems. It can treat very large systems with
hundreds of spins or electrons, and it is extremely accurate for 
quasi one-dimensional systems. 
The basic idea consists of truncating the Hilbert space in such a way
as to maximize the amount of information we preserve about the
actual ground state of the  system.
In order to do this, we need to systematically apply a change of basis 
called the density matrix projection. The resulting wavefunction is a very 
accurate representation of the ground-state in a reduced Hilbert space 
with a given number of states, and assumes the form of a matrix product state.
In this paper we want to apply the DMRG to study a two-dimensional 
system of electrons in a strong a magnetic field. 
The generic form of the Hamiltonian is 

\begin{eqnarray}
H=\frac{1}{2}\sum_{\substack{m_1,m_2\\m_3,m_4}} \langle m_1,m_2|V|m_3,m_4 \rangle c^{\dagger}_{m_1} c^{\dagger}_{m_2} c_{m_4} c_{m_3}
\label{hami}
\end{eqnarray}
where the indices $m$ represent the quantum numbers of free electron 
orbitals in a given Landau level (LL), and $V$ is the electron-electron
interaction (which will simply be the Coulomb repulsion in this paper
although one could modify the interaction in order to account for
finite layer thickness, for instance). 
Energies are quoted in units of $e^2/4\pi\epsilon \ell_0$,
where $\ell_0=(\hbar c/eB)^{1/2}$ is the magnetic length. The orbitals on a spherical 
geometry are labeled by their angular momentum, while on the torus, one 
needs to use the magnetic translations, identified by the two components 
$k_x$, and $k_y$. We have omitted the spin degrees of freedom due to the 
large Zeeman splitting. Thus, this Hamiltonian represents a 
two-dimensional system of spinless electrons with long-range interactions. 
In order to apply DMRG, one can use a generalization of the algorithm to 
momentum space, which has already been applied to the Hubbard 
model\cite{kdmrg}. 
We note that the extension of the DMRG to $2D$ systems
is an important challenge which is a subject of great current interest \cite{uli}.
Our method exploits the Landau level structure of the FQH problem
by applying the momentum space formulation of the DMRG.

\begin{centering}
\begin{figure}
\epsfig {file=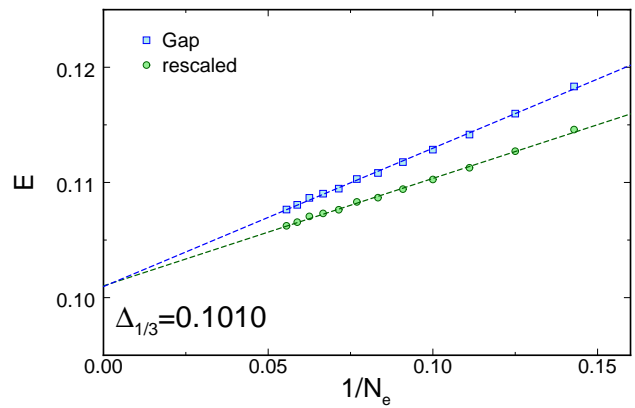,width=55mm, angle=-90}
\caption{
Excitation energy at $\nu=1/3$ obtained with DMRG and exact diagonalization, and the extrapolation using a linear function of $1/N_e$. We also show the results for the rescaled energies using a size-independent magnetic length.
}
\label{fig2}
\end{figure}
\end{centering}

The idea is to arrange the orbitals along a one-dimensional chain, and 
apply the finite-system version of the DMRG.
The simulation starts with a small system with a few orbitals that can be 
easily diagonalized. We split the system in two halves: the left block 
($B_L$) and the right block ($B_R$). We increase the system size by 
systematically adding orbitals to the system. We do that by including two 
orbitals in the center, between the blocks, in a configuration $B_L 
\bullet \bullet B_R$. We then calculate the ground state of the 
Hamiltonian, and obtain the reduced density matrix of one block plus the 
new added site, say $[B_L \bullet]$, by tracing over the basis states of 
$[\bullet B_R]$. We pick the new basis for the new block $B_L$ as the 
eigenstates of the reduced density matrix with the largest eigenvalues.
We can adjust how many basis states we keep, thereby controlling
the accuracy of the computation. Once the desired system size has 
been reached, we grow one block and shrink the other one, sweeping from 
left to right, and right to left, until we converge. The momentum space 
version of the DMRG allows us to use the total momentum (angular momentum 
on the sphere), besides the particle number $N_e$, to work in a subspace
of reduced dimension. The matrix elements of the interaction $V$ on the sphere take the form 
\cite{HR1,Fano} 
\begin{multline}
\langle m_1,m_2|V|m_3,m_4 \rangle =  \\
\sum_{J=0}^{2S} \sum_{M=-J}^{+J} \langle S,m_1,S,m_2|JM\rangle \langle JM|S,m_3,S,m_4 \rangle V_J/R, 
\end{multline}
where $R$ is the radius of the sphere, and the coefficients $V_J$ are 
Haldane's pseudopotentials \cite{Haldane}, derived from the pure Coulomb potential. 
They represent the interaction 
energy of two electrons with relative angular momentum $J$, and depend on 
the Landau level. To obtain this expression, we have used addition of 
angular momenta, and $2S = N_\Phi$ is the total flux through the system
(not to be confused with electron spins, which are assumed to be fully
polarized), given by the number of electrons $N_e$ and the filling fraction $\nu$. As shown by 
Haldane \cite{Haldane}, the Laughlin state at $\nu=1/3$ is the exact ground state for a potential
obtained by setting all ${V_J}=0$ except for $V_1$. We use this observation
to compute the Laughlin state in the same basis as the ground state
of the system for Coulomb interactions. 
In order to represent both states in the same basis simultaneously
with equivalent accuracy we calculate the ground states of the full Coulomb interaction, 
$\psi$, and Haldane's Hamiltonian, $\psi_0$, and we use both states to 
obtain the reduced density matrix at each DMRG step \cite{dmrg}.
This is essential for us to estimate variational energies
as well as overlaps between the Laughlin state and the exact ground state of the system.

\begin{centering}
\begin{figure}
\epsfig {file=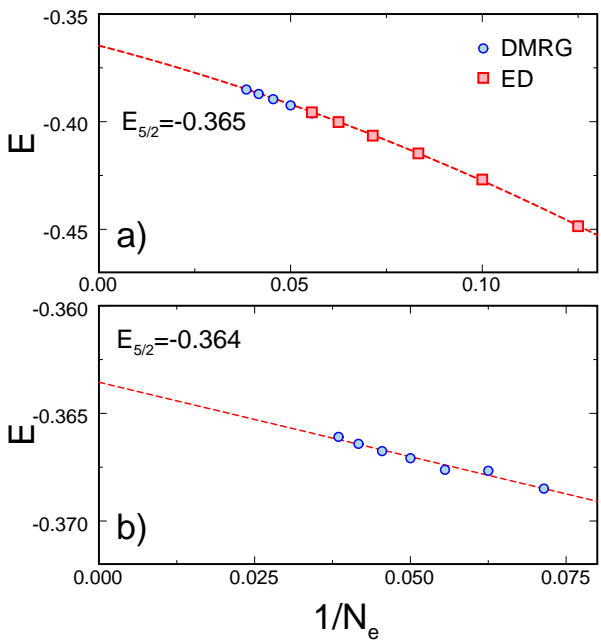,width=80mm}
\caption{
a) Ground state energies at $\nu=5/2$ for DMRG and exact diagonalization. 
b) Finite size scaling using a renormalized magnetic length $\ell_0$, following Refs.\cite{Morf, Morf and Shankar}.
}
\label{fig3}
\end{figure}
\end{centering}

\paragraph{Optimization strategies --}
Since the Laughlin and Moore-Read states have non-zero
amplitude only on ``squeezable'' configurations \cite{squeezing},
we have replaced the traditional DMRG
warmup scheme by starting our simulation from a small set of
``squeezable'' states, thereby improving convergence considerably.
This also removes undesired effects, such as sticking to an excited state. 
Another important ingredient of our algorithm is the  
addition of density matrix corrections \cite{dm corrections} 
that takes into account the incompleteness of the environment block and the 
missing entanglement due to the non-locality of the interactions. These 
corrections improve the energies by one order of magnitude compared to the
conventional algorithm. These optimizations have allowed us to 
obtain ground state energies and gaps for systems with up to $N_e=20$ electrons
at $\nu=1/3$ and $N_e=26$ at $\nu=5/2$, using up to $5000$ DMRG states, with a typical truncation error of the order of $10^{-6}$ for the largest systems studied here. It is interesting to note that the size of the DMRG Hilbert space is of the order of $1,000,000$ states, while the full Hilbert space has a total dimension of the order of many millions for $N_e \geq 12$.

\begin{table}[t]
\begin{tabular}{c|c|c|c}
 & DMRG & Variational & Overlap \\
$N_e$ & $E(\frac{1}{3})$ & $E_0$ & $\langle \psi | \psi_0 \rangle$ \\ \hline 

12 & -0.428801 & -0.428521 & 0.99428 \\
13 & -0.427278 & -0.426989 & 0.99383 \\
14 & -0.425982 & -0.425689 & 0.99382 \\
15 & -0.424871 & -0.424571 & 0.99316 \\
16 & -0.423903 & -0.423597 & 0.99332 \\
17 & -0.423051 & -0.422745 & 0.99253 \\
18 & -0.422299 & -0.421995 & 0.99312 \\
19 & -0.421621 & -0.421319 & 0.9930 \\
20 & -0.420998 & -0.420720 & 0.99290 \\

\end{tabular}
\caption{
Ground state energies $E_{1/3}$ and variational energies $E_0$ for $\nu=1/3$ obtained with DMRG. The last column shows the overlap between the exact wave function $\psi$ and the variational (Laughlin) state $\psi_0$ obtained using Haldane's Hamiltonian.
}
\label{overlaps}
\end{table}

\paragraph{Results --}

In Table I, we show the variational energy of the Laughlin state compared to the actual ground state energy of the full Coulomb Hamiltonian at $\nu=1/3$, at flux $N_\Phi=3(N-1)$. These were obtained using exact diagonalization for $N \leq 12$,\cite{Morf communication} and DMRG for larger systems. We also show the overlap between both wavefunctions, indicating unambiguously that the Laughlin state does, indeed, describe the physics of the
$\nu=1/3$ FQHE as we approach the thermodynamic limit.
The energies are also plotted in Fig.\ref{fig1}. Extrapolations to the thermodynamic
limit can be carried out using a second-order polynomial in $1/N_e$, see Fig. \ref{fig1}(a), or a first-order
one after renormalizing the magnetic length $\ell _0$ to take into account the curvature of the sphere,
which puts all the data points on a straight line \cite{Morf and Halperin, Morf, Morf and Shankar}.
Notice that the scale shown in the y-axis in Fig.\ref{fig1}(b) is much smaller than the original,
making the DMRG errors visible for large systems. However these errors are small,
of the order of $10^{-5}$, and one order of magnitude smaller for the variational energy
of the Laughlin state.\cite{footnote}
It can clearly be seen that finite-size effects are very small for $N \geq 10$. The fact that accuracy and convergence are better for the Laughlin wavefunction can be attributed to the short range of the interactions and the fact that it spans a reduced Hilbert space \cite{squeezing},
allowing DMRG to efficiently represent it as a matrix product state \cite{laughlin MPS}.
The extrapolated values of $E_0=-0.40977$ and $E_{1/3}=-0.41010$ for the Laughlin state and exact ground state energies, respectively, agree well with previous Monte Carlo \cite{Morf and Halperin} and exact diagonalization\cite{Morf} estimates of $E_0=-0.40975$, and $E_{1/3}=-0.4101$.

In Fig.\ref{fig2} we show DMRG results for the energy gap at $\nu=1/3$, calculated from the energy of charged excitations with flux $N_\Phi = 3(N-1)\pm 1$. Our extrapolated value in the thermodynamic limit of $\Delta_{1/3}=0.1010$ agrees well with previous estimates of $\Delta_{1/3}=0.1012$ using smaller systems up to $N_e=12$. \cite{Morf and Shankar}

Having successfully tested the DMRG at $\nu=1/3$, we proceed by extending the analysis to the second Landau level, at $\nu=5/2$. In Fig.\ref{fig3}(a) and (b), we show the ground state energies at flux $N_\Phi=2N-3$, and the scaled corrected values using a size-independent magnetic length, respectively. Extrapolations yield a value of
$E_{5/2}=-0.364$, which compares well with Morf's result \cite{Morf} of $-0.366$. Notice that in our study we have ignored system sizes smaller that $N_e=10$ because of strong finite-size effects.

\begin{centering}
\begin{figure}
\epsfig {file=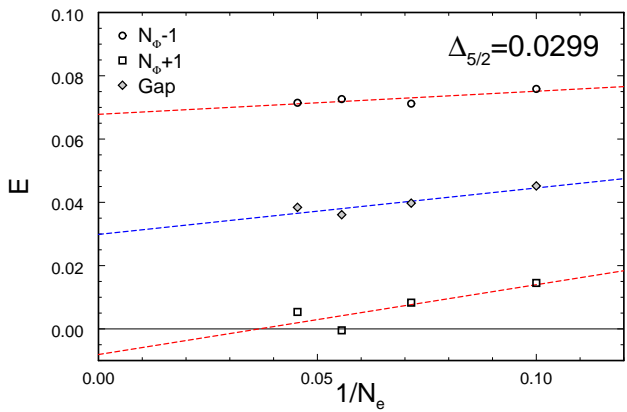,width=55mm,angle=-90}
\caption{
Size dependence of the excitation energies and gap at $\nu=5/2$, obtained with DMRG.
}
\label{fig4}
\end{figure}
\end{centering}

The calculation of the gap at $\nu=5/2$ presents an extra complication due to `aliasing' with states corresponding to conventional FQH fractions \cite{Morf and Shankar, Morf}. This constrains the number of accessible states to the sequence $N_e = 10,14,18,...$. Due to numerical limitations, we have been able to add a single extra point to this sequence with $N_e=22$. The results are depicted in Fig.\ref{fig4}, 
where we show the energies for the half-flux-quantum $(e/4)$ quasiparticle and quasihole, in addition to the gap, which is their average. As may be seen from the figure there is more scatter in the quasihole and quasiparticle energies than in the gap. In particular, the energies at $N_e=18$ look anomalous. This suggests strong finites-size effects, and an alternative extrapolation based on only the two largest system sizes would give a significantly larger extrapolated gap value. Details will be presented elsewhere. The results for smaller system sizes coincide with Morf's \cite{Morf, Morf and Shankar}, and a linear extrapolation using all the datapoints yields a value of the gap of $\Delta_{5/2}=0.030$, also larger than the result of $\Delta_{5/2}=0.025$ previously reported. 
We emphasize that our estimated gap is an upper bound to the real gap since, following Morf \cite{Morf,Morf and Shankar}, we treat the quasiholes as point particles.

\paragraph{Conclusion --}
We have developed and validated the DMRG technique as a theoretical tool 
for the numerical study of FQH incompressible states, thereby 
increasing the possible tractable system size $N_e$ by a 
nearly a factor 2 (e.g. from $N_e = 10$ to $N_e = 20$ electrons for $\nu = 1/3$).  
We have applied the technique to $\nu = 1/3$ and $5/2$, obtaining in the process
the best estimates for both the ground state energy and the excitation gap for 
these important FQH states. We have also significantly expanded the 
regime of validity for the infinite system extrapolation, $1/N_e \rightarrow 0$, 
which is essential for any comparison with experimental work.  
We have found that a reliable finite size scaling requires a large number of DMRG states (up to 5000 in our work), and that density matrix corrections and "squeezing" are essential to improve the accuracy and convergence. Furthermore, this technique can be used to calculate and validate variational states, such as the Laughlin state. 
Our work 
establishes the DMRG as a numerical tool for future theoretical FQHE studies 
where large values of $N_e$ are essential for theoretical understanding.
In particular, the DMRG technique 
may be able to to shed light on various mysterious fractions such
as $5/11$, $6/13$, $4/13$, $3/8$, and
should be useful in studying incompressible states
in higher LLs, 
where large system sizes are essential in distinguishing between
competing compressible and incompressible phases \cite{Rezayi00} which lie
close in energy.

\acknowledgements
AEF thanks S.R. White and V. Scarola for useful discussions.
We are very grateful to R. Morf for a careful reading of the manuscript and insightful comments.
This research has been supported by Microsoft Station Q,
the NSF under DMR-0411800 (CN),
and DMR-0606566 (EHR), and by the ARO under grant W911NF-04-1-0236 (CN and SDS).


\begin{thebibliography}{99}
\bibitem{footnote} Errors in the variational energy of the Laughlin state are due to the
fact that we compute it by applying the DMRG algorithm to the $V_1$-only Hamiltonian
rather than by expanding its analytic form and transforming it into the truncated basis for
the exact ground state. This also tends to systematically overestimate the
overlaps shown in Table \ref{overlaps} slightly.
\end{thebibliography}
\end{document}